\documentstyle[12pt,equation]{article}
\begin{document}
\begin{titlepage}
\begin{flushright}
TAUP-2241/95\\
hep-ph/9503305\\
March 1995
\end{flushright}
\vskip 2cm
\begin{center}
{\Large \bf Charmonium in a weakly coupled quark-gluon plasma
\\ \ \\}
Daphne Levin-Plotnik\footnote{E-mail: daphne@albert.tau.ac.il}\\
Benjamin Svetitsky\footnote{E-mail: bqs@albert.tau.ac.il}\\
\ \\
{\it School of Physics and Astronomy\\
Raymond and Beverly Sackler Faculty\\
of Exact Sciences\\
Tel Aviv University\\
69978 Tel Aviv, Israel}
\end{center}
\vskip 2cm
\noindent
{\it ABSTRACT.\/}  We present a model of charmonium as two heavy
quarks propagating classically in a weakly coupled quark-gluon plasma.
The quarks interact via a static, color-dependent potential and
also suffer collisions with the plasma particles.
We calculate the radiation width of the color octet state
(for fixed, classical $q\bar q$ separation)
and find that it is long-lived provided
a finite gluon mass is used to provide a threshold energy.

\end{titlepage}
Theoretical studies of quarkonium in the quark-gluon plasma have
generally treated the $q\bar q$ bound state as a unit, a color-singlet
meson propagating in the medium \cite{Petronzio,Blaizot1,Ruuskanen,Chu},
evolving
slowly \cite{Horvath,Cleymans,Cerny,Prorok},
and colliding with the plasma's constituents
\cite{Zahed,Ropke}.
Paradoxically, the interaction binding the $q\bar q$ is taken to be
a static, screened Coulomb potential \cite{Matsui,Chu2}, which implies that the
density
of plasma particles is high enough to allow construction of such
a smooth potential---a weakly coupled plasma \cite{Ichimaru,Hosoya}.
In this note we adopt the second point of view and construct
a model accordingly.
We consider \cite{thesis} a picture of quarkonium
as two heavy particles propagating classically, subject to a screened
potential;
these particles collide {\em individually\/} with plasma particles,
changing momentum and color thereby.
Such a model can be used to study in detail the diffusion of unbound
$c\bar c$ pairs and the formation and destruction of charmonium
in the plasma.

For simplicity, we specify the potential in the plasma rest frame, and
neglect the associated color-magnetic interaction.
In writing the $q\bar q$ interaction we keep track of the color state
$c$ of the pair.
The potential takes the form
\begin{equation}
V_c(r)=\zeta_c \frac{\alpha_s}r \exp(-r/r_{_D})\ ,
\label{Debye}
\end{equation}
where $\zeta_c=-4/3$~and~$1/6$ for the singlet and octet states,
respectively.
We distinguish {\em a priori\/} three regimes in $r$ (see Fig.~1):
\begin{enumerate}
\item
For large $r$ the interaction is negligible and the system is
color-blind---the two quarks diffuse independently in the plasma \cite{bqs},
and their color state (which is irrelevant to their motion) is
determined statistically.
\item
For intermediate $r$ one must keep track of the color state in detail,
and propagate the quarks according to the potential appropriate to
the instantaneous color state.
\item
For small $r$ the large gap between singlet and octet might induce
a large transition rate due to gluon radiation.
If the width $\Gamma_8$ of the octet state is larger than the gap
$\Delta V$, the octet ceases to exist as an excited state of the pair,
vanishing into the $(q\bar q)_{\rm singlet}+ g$ continuum.
\end{enumerate}
The boundaries of these regions must be determined by explicit calculation
of transition rates.

Before presenting the formal kinetic theory, we address the
question of the existence of Region 3.
Transitions between the color states are caused by collisions of the
individual heavy particles with plasma particles, and by radiation and
absorption of gluons.
Only radiation contributes
to the octet state's width in the lowest order in $\alpha_s$.
The radiating quark moves in the potential field of its partner, which
is otherwise a spectator to the process (see Fig.~2).
For an initial quark momentum $p$ relative to the plasma,
the radiation rate is given by
\begin{eqnarray}
\Gamma(p)&=&\frac1{2E_p}\int \frac{d^3p'}{(2\pi)^3 2E_{p'}}
\frac{d^3k}{(2\pi)^3 2k}
\tilde g(k)(2\pi)^4 \delta^3({\bf p'}+{\bf k}-{\bf p})\nonumber\\
&&\qquad\qquad\qquad\times\,
\delta\bigl(k+E_{p'}+V_1(r)-E_p-V_8(r)\bigr)
\sum \left|{\cal M}\right|^2\,.
\label{rate}
\end{eqnarray}
Here $\tilde g(k)$ is the statistical factor for the outgoing gluon;
the matrix element squared is averaged over initial spin and summed
over final spins of the gluon and the participating quark.
The summation over colors must be done carefully because of the condition
that the initial state be an octet and the final state, a singlet.
We define a basis of color state vectors $|ij\rangle$, where $i$,~$j=1$,~2,~3
are the color indices of the quark and antiquark, respectively.
Then the singlet state is
\begin{equation}
|1\rangle=\frac1{\sqrt3}\sum_i |ii\rangle
\label{singlet}
\end{equation}
while a member of the octet is\footnote{We normalize the color matrices so that
$\hbox{Tr}\,T^aT^b=\frac12\delta^{ab}$.}
\begin{equation}
|8r\rangle=\sqrt2 \sum_{ij}T^r_{ij}|ij\rangle\ .
\label{octet}
\end{equation}
The Feynman matrix element then contains the color factor
\begin{equation}
C^{ar}=\langle1|T^a|8r\rangle=\frac1{\sqrt6}\delta^{ar}\ ,
\end{equation}
where the color operator $T^a$ acts on either the quark or
the antiquark state only.
Squaring the matrix element, summing over the color $a$ of the emitted
gluon, and averaging over $r$ gives
\begin{equation}
\frac18\sum_{a,r}|C^{ar}|^2=\frac16
\end{equation}
The result is
\begin{equation}
\sum \left|{\cal M}\right|^2=\frac{8\pi\alpha_s}3\left(2m^2-p\cdot p'\right)\ .
\end{equation}
Of course, either the quark or the antiquark may radiate;
if their momenta are $p_1$ and $p_2$, then the total radiation
rate from the pair is $\Gamma_8=\Gamma(p_1)+\Gamma(p_2)$.

Calculating $\Gamma(p)$ from (\ref{rate}) gives the nonsensical result
that $\Gamma(p)\neq0$ in the limit $\Delta V\to0$.
In this limit
\begin{equation}
\Gamma(p=0)\sim \frac{\alpha_s}{3m}\int \frac{k\,dk}{k+\sqrt{k^2+m^2}}
\sum \left|{\cal M}\right|^2 \tilde g(k)\delta(k-\Delta V)\ .
\end{equation}
The source of the difficulty is that the Bose-Einstein factor contributes
$1/k$ to counteract the vanishing phase space \cite{Milana}.
It is evidently a bad approximation to take the radiated gluon to
be massless;
the mass induced by the plasma must be taken into account.
The simplest thing to do is to take (\ref{rate}) and to
replace $k$ in the measure, in $\tilde g(k)$, and in the
energy $\delta$ function by $\sqrt{k^2+m_g^2}$.
This makes $\Gamma$ vanish for gap values below the threshold
$\Delta V=m_g$, as shown in Fig.~3.
A more correct calculation would replace the energy $\delta$ function
by the one-loop spectral density of transverse gluons \cite{Braaten,Blaizot2},
and include as well the radiation of longitudinal plasmons, similarly
treated.

The qualitative result indicated by Fig.~3 is that there is no
distinct small-$r$ regime as we supposed.
The octet state is meaningful
even when $\Delta V$ is large.
Of course, $\Gamma_8$ will be increased by the effects of collisions,
and so it is still possible that the octet disappears at some
value of $r$.

The radiation and collision rates are
ingredients in the Boltzmann-Vlasov equation which governs the
evolution of the $q\bar q$ distribution.
This distribution is defined as a function
$f_c({\bf x}_1,{\bf x}_2,{\bf p}_1,{\bf p}_2;t)$ on the two-particle
phase space, carrying as well the index $c$ denoting the color state
of the pair.
It satisfies the transport equation
\begin{equation}
\frac D{Dt}f_c=\left(\frac{\partial f_c}{\partial t}\right)_{\rm collision}
\ .
\label{BV}
\end{equation}
The left-hand side is the convective derivative in phase space,
\begin{equation}
\frac D{Dt}f_c\equiv \left[\frac{\partial}{\partial t}+
\frac{{\bf p}_1}m \cdot \frac{\partial}{\partial {\bf x}_1}+
\frac{{\bf p}_2}m \cdot \frac{\partial}{\partial {\bf x}_2}+
{\bf F}_c({\bf x}_1-{\bf x}_2)\cdot
\left(\frac{\partial}{\partial {\bf p}_1}-\frac{\partial}{\partial {\bf p}_2}
\right)\right]f_c\  ,
\label{LHS}
\end{equation}
where ${\bf F}_c({\bf r})=-\nabla V_c(r)$ is the color force.
The right-hand side of (\ref{BV})
represents gain and loss through collisions and radiation.
In the small- and intermediate-distance regimes, we distinguish
collisions which change the color state from those which do not.
The former (like the radiation process)
carry momentum-transfer of at least $\Delta V(r)$,
and hence are hard collisions; the latter may carry very small
momentum transfer, and hence may be approximated as soft.
Thus we write
\begin{equation}
\left(\frac{\partial f_c}{\partial t}\right)_{\rm collision}=
R_{\rm soft}+R_{\rm hard}\ .
\end{equation}
We can approximate the soft collision term by Fokker-Planck
terms \cite{bqs},
\begin{equation}
R_{\rm soft}=-\frac{\partial}{\partial {\bf p}_1}
\cdot[{\bf G}_c({\bf p}_1)f_c]
+\frac{\partial^2}{\partial p_{1i}\partial p_{1j}}
[N^c_{ij}({\bf p}_1)\,f_c]+
(1\to2)\ .
\end{equation}
${\bf G}_c$ and $N^c_{ij}$ are drag and diffusion coefficients,
calculated in the Landau approximation as moments of transition
rates \cite{Landau,bqs}.
Here, as for the radiation rate calculated above, one has to be careful
in projecting out the proper color channel.
This makes ${\bf G}_c$ and $N^c_{ij}$ dependent on $c$, and
implies that only a subset of the scattering diagrams contributes
to a given color channel.

We have calculated the drag and diffusion coefficients
in the two color channels \cite{thesis}.
These can come only from soft collision processes, since radiation
necessarily flips color.
The octet coefficients are similar to
those for the color-blind scattering
which a single quark undergoes \cite{bqs}, reduced by roughly
a factor of 2--3 because of the restriction of the final states.
The coefficients for the singlet state are smaller by an additional
factor of 100, again because of the
reduced number of final states, and also because of the smaller
set of diagrams which contribute.

The hard collision terms cannot be approximated, and must be written
as full-blown collision integrals.
We have the loss and gain terms,
\begin{eqnarray}
R_{\rm hard}=&-&\left[\Gamma_c({\bf p}_1)+\Gamma_c({\bf p}_2)\right]
f_c({\bf x}_a,{\bf p}_a;t)\nonumber\\
&+&\int d^3k\,w({\bf p}_1-{\bf k}\to{\bf p}_1\,;\,\bar c\to c)\,f_{\bar c}
({\bf x}_a,{\bf p}_1-{\bf k},{\bf p}_2;t)\nonumber\\
&+&\int d^3k\,w({\bf p}_2-{\bf k}\to{\bf p}_2\,;\,\bar c\to c)\,f_{\bar c}
({\bf x}_a,{\bf p}_1,{\bf p}_2-{\bf k};t)\ .
\end{eqnarray}
Here $\Gamma_c({\bf p})$ is the total rate of radiation- and
collision-induced transitions for each quark, along the lines
of $\Gamma$ defined in (\ref{rate}).
The functions $w$ are differential transition rates into the
phase space element at $({\bf x}_a,{\bf p}_a)$ from other
elements;
since these are hard collisions, they involve perforce a
color flip into $c$ from the complementary state $\bar c$,
and depend on position through the value of $\Delta V(r)$.

Our result for the radiative width of the octet state gives
a lifetime of several fm/$c$ for reasonable
$q\bar q$ separations.
If the collision rates do not affect this result strongly,
then the evolution of the pair in phase space
requires detailed numerical solution of the Boltzmann
equation.
If, on the other hand, the collisional width is large,
then it may be reasonable to assume that the singlet
and octet states are in equilibrium.
Then a simplified Boltzmann equation may be studied, in
which $f_c\propto\exp(-V_c/T)$.
If the collisional width turns out to exceed the gap $\Delta V$
at some values of $r$,
then the octet state ceases to exist
and we are back to the consideration of three distinct
regimes in $r$ as described above.

\section*{Acknowledgements}
This work grew out of conversations with Tetsuo Matsui.
We thank Judah Eisenberg for fruitful discussions.
This work was supported by a Wolfson Research
Award administered by the Israel Academy of Sciences and
Humanities, and by the U.S.-Israel Binational Science Foundation
under Grant No.~92-00011.

\end{document}